\documentclass[preprint,12pt]{elsarticle}

\usepackage{siunitx}
\usepackage{rotating}

\usepackage[pdftex]{}
\usepackage[pdftex]{color}
\usepackage{float}
\usepackage{amsmath,amssymb}

\journal{Preprint submitted to Nucl. Instr. Meth. A}

\begin{document}
\begin{frontmatter}



\title{Development of a low-$\alpha$-emitting $\mu$-PIC as a readout device for direction-sensitive dark matter detectors}


\author[aff1]{Takashi Hashimoto}
\author[aff1]{Kentaro Miuchi}
\author[aff1]{Tomonori Ikeda}
\author[aff1]{Hirohisa Ishiura}
\author[aff1]{Kiseki D. Nakamura}
\author[aff1]{Hiroshi Ito}
\author[aff2,aff3]{Koichi Ichimura}
\author[aff2,aff3]{Ko Abe}
\author[aff2,aff3]{Kazuyoshi Kobayashi}
\author[aff4]{Atsushi Takada}
\author[aff1]{Atsuhiko Ochi}
\author[aff1]{Takuma Nakamura}
\author[aff1]{Takuya Shimada}
\address[aff1]{Department of Physics, Kobe University, Kobe, Hyogo 657-8501, Japan}
\address[aff2]{Kavli Institute for the Physics and Mathematics of the Universe (WPI), the University of Tokyo, Kashiwa, Chiba, 277-8582, Japan}
\address[aff3]{Kamioka Observatory, Institute for Cosmic Ray Research, the University of Tokyo, Higashi-Mozumi, Kamioka, Hida, Gifu, 506-1205, Japan}
\address[aff4]{Kyoto University, Kitashirakawaoiwake-cho Sakyo-ku Kyoto-shi Kyoto, 606-8502, Japan}

\begin{abstract}
\par
Direction sensitivity could provide robust evidence for the direct detection of weakly interacting massive particles constituting dark matter.
However, the sensitivity of this method remains low due to the radioactive backgrounds.
The purpose of this study is to develop a low-background detector as a two-dimensional imaging device for a gaseous time projection chamber.
In direction-sensitive dark matter experiments~(e.g. NEWAGE), $\alpha$-rays emitted from the detector components often create substantial radioactive backgrounds.
Based on the study of the background of NEWAGE, a new detector ``low-$\alpha$ $\mu$-PIC'' is developed.
The produced $\mu$-PIC performs well as a gas detector and the $\alpha$-ray emission rate from the $\mu$-PIC reduced by a factor of 100.
\end{abstract}

\begin{keyword}
Gaseous detector,
Micro-pattern detector,
$\mu$-PIC




\end{keyword}

\end{frontmatter}


\section{Introduction}
\label{}
\begin{figure}[H]
\centering
\includegraphics[width=10cm]{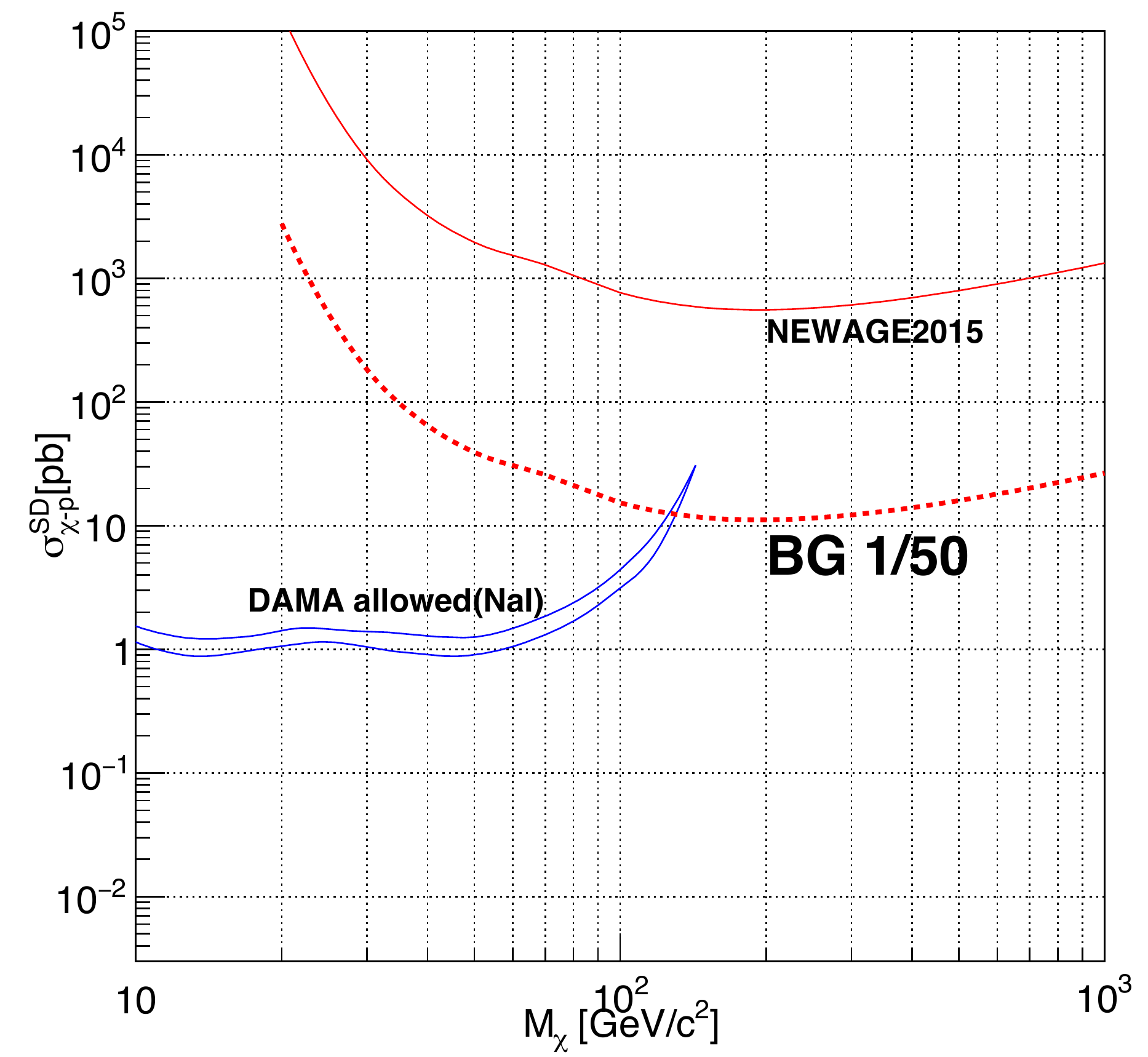}
\caption{Latest result of NEWAGE~(``NEWAGE2015'')~\cite{NEWAGE_2015} and expected search sensitivity with a background rate reduced by a factor of 50~(``BG~1/50'').
``DAMA allowed (NaI)'' is the region where the WIMP is said to exist as an interpretation of the DAMA results~\cite{DAMA_2004}.
The horizontal axis shows the mass of the WIMPs, and the vertical axis shows the scattering cross section of the spin-dependent~(SD) interaction of protons and WIMPs.
} 
\label{fg:future_limit_SD}
\end{figure}

A large fraction (\textasciitilde26\%). of the universe is in the form of non-baryonic cold dark matter~\cite{Planck}.
Weakly interacting massive particles (WIMPs) are possible dark matter candidate particles~\cite{PWIMP}.
Despite numerous experimental efforts, WIMPs have still not been observed directly, except for the signal reported by the DAMA/LIBRA~\cite{DAMA_2018}, CRESST~\cite{CRESST} and CDMS~\cite{CDMS} experiments.
However, only DAMA/LIBRA experiment continues to observe and claim.
On the other hand, several other experiments have reported contradictory results, and further investigation is required~\cite{XENON,LUX,PANDA}.
Therefore, none of positive detection signatures of WIMPs has been universally accepted.
Direction-sensitive methods have been suggested to provide signatures that are more convincing for WIMPs~\cite{DM_direction_strong_evidence}.
Because the Cygnus constellation is seen in the traveling direction of the Solar System through the galaxy, any galactic-halo WIMPs would appear to originate from the Cygnus direction as a ``WIMP-wind".
Therefore an asymmetric distribution of incoming WIMPs would provide strong evidence of WIMP detection. 
Direction-sensitive WIMP search experiments are often designed to measure both the energy and track of a recoil nucleus.
Among several methods being developed intensively around the world~\cite{NIT_NIM2007,CNT_DM}, a gaseous time projection chamber using a low-pressure gas is considered most promising and has been studied for decades~\cite{DRIFT_APP2012,DMTPC_PLB2011,MIMAC_2011}.
NEWAGE is a direction-sensitive dark-matter search experiment using a gaseous three-dimensional tracking detector, or a micro-time projection chamber ($\mu$-TPC).
NEWAGE-0.3b', one of the NEWAGE $\mu$-TPC detectors, comprises a two-dimensional fine-pitch imaging device called a micro-pixel chamber ($\mu$-PIC) ~\cite{uPIC}; a gaseous electron multiplier (GEM)~\cite{GEM}; and a detection volume (30$\times$30$\times$41~cm$^{3}$) filled with  CF$_4$ gas at 0.1~atm.
The area of the $\mu$-PIC is $30.7\times30.7~\si{cm^{2}}$ read by 768-anode and 768-cathode strips with a pitch of 400~\si{\micro \metre}.
NEWAGE has conducted direction sensitive dark-matter search experiments in the Kamioka underground laboratory since 2007 and set the direction-sensitive SD cross-section limit of 557\,\si{pb} (at 90$\%$ confidence level) for a WIMP mass of 200\,$\si{GeV/}c^{2}$~\cite{NEWAGE_2015}.
The detector performance and direction-sensitive limits of the NEWAGE-0.3b' were described in Ref.~\cite{NEWAGE_2015}.
The ultimate goal of the directional method is to explore even beyond the neutrino floor , which will require large and extremely low background detectors\cite{DDSEARCH1,DDSEARCH2}. In the meanwhile, since the existing three-dimensional tracking detector which shows the best directional limits is our $\mu$-TPC, the DAMA region is set as the next realistic milestone (``DAMA allowed (NaI)'' in Figure~\ref{fg:future_limit_SD}).
One of the expected results of NEWAGE with sensitivity improvements by a factor 50 is illustrated in Figure~\ref{fg:future_limit_SD}. 
NEWAGE can start the investigation of the DAMA region with this improvement.
The main background in NEWAGE2015 was found to be the $\alpha$-ray emissions from the $\mu$-PIC.
When the $\alpha$-ray emission rate becomes 1/100, the detection sensitivity is expected to improve 100 times.
However, estimating the $\alpha$-ray background near the threshold has ambiguities due to some uncertainties in the $\alpha$-ray background estimation. So we have the margin of a factor~2.

\section{Development of Low-$\alpha$ $\mu$-PIC}
\subsection{Background study for NEWAGE-0.3b'}
\begin{figure}[ht]
		 \centerline{\includegraphics[width=150mm]{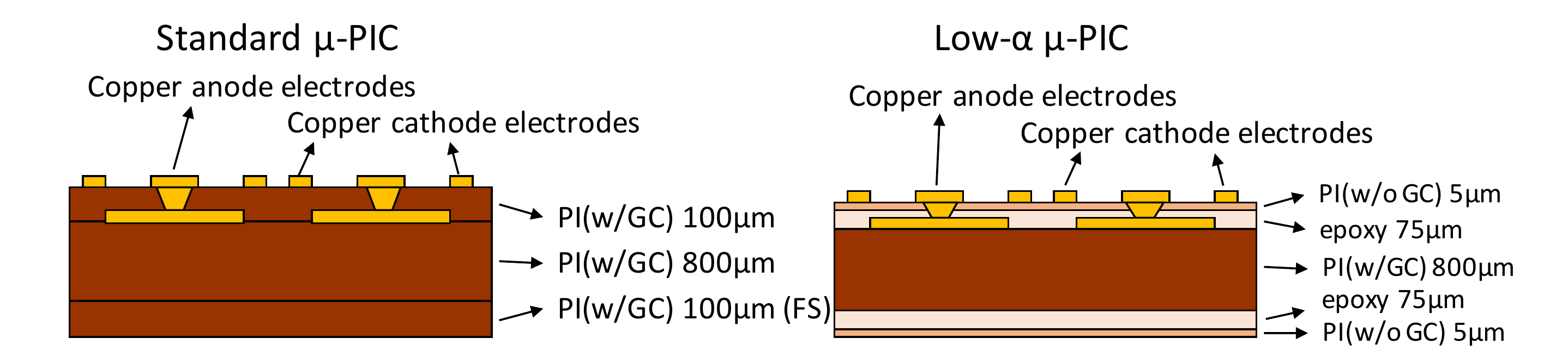}}
  \caption{Cross section view of the standard $\mu$-PIC (left) and low-$\alpha$ $\mu$-PIC (right).}
  \label{fig:cs_uPIC}
\end{figure}

To improve the sensitivity of the NEWAGE-0.3b', it is necessary to understand and reduce the background events. 
Radioactive contaminants in the detector components are well-known background sources in rare-event searches.
In particular, many materials contain the natural radioactive isotopes $^{238}$U and $^{232}$Th (U/Th)\cite{BG}.
These isotopes emit several $\alpha$-, $\beta$- and $\gamma$-rays in their decay series.
Simulation studies indicated that $\alpha$-rays from the U/Th series contained in the $\mu$-PIC were the main source of background~\cite{D_nakamura}.
As discussed in Ref.\cite{D_nakamura}, studies show that other backgrounds (e.g. ambient neutron and $\gamma$-ray) are considered not to become the main background even after reducing $\alpha$-rays background by a factor of 100.
The U/Th contamination of the detector components around the $\mu$-PIC was measured using a high-purity germanium (HPGe) detector.
The $\mu$-PIC used in the previous dark matter search experiments~(``NEWAGE2015''~\cite{NEWAGE_2015}) is referred to as ``standard $\mu$-PIC'' hereafter.
A standard $\mu$-PIC consists of three layers as shown in Figure~\ref{fig:cs_uPIC}~(left) : a layer of polyimide reinforced with a glass cloth-sheet~PI(w/GC) 100~$\si{\micro\meter}$ part, PI(w/GC) 800~$\si{\micro\meter}$ part and another PI(w/GC) 100~$\si{\micro\meter}$ part.
The radioactivities of the PI(w/GC) 100~$\si{\micro\meter}$ layer and the PI(w/GC) 800~$\si{\micro\meter}$ layer were independently measured. 
The radioactivity of the glass sheet taken out of the PI(w/GC) 100um layer was also measured.
The surface plating solution~(CuSO$_{4}$) used for $\si{\micro}$-PIC and the GEM that was used close to the detection volume were also measured.
Radioactivities of the upper and middle streams of the U series (${}^{238}\rm{U}$ to ${}^{226}\rm{Ra}$ and ${}^{226}\rm{Ra}$ to ${}^{210}\rm{Pb}$, respectively) and the Th series were obtained from the peak of 93~keV by the ${}^{234}\rm{Th}$ decay, the peak of 609~keV by the decay of the ${}^{214}\rm{Bi}$, and the peak of 583~keV by the decay of the ${}^{208}\rm{Tl}$, respectively.
The radioactivities of the middle stream of the U series were converted to ${}^{238}\rm{U}$-equivalent contamination in order to examine the radiation equilibrium.
\par
The measurement results are listed in Table~\ref{tab:measurment_result}.
Since the finite values were not obtained for ${}^{238}\rm{U}$ and $^{232}\rm{Th}$ contents of the GEM and the plating solution, 90\% confidence level upper limits were set.
The PI(w/GC) 800~$\si{\micro\meter}$ and 100~$\si{\micro\meter}$ parts and the glass cloth contained in the PI(w/GC) 100~$\si{\micro\meter}$ part had finite values.
It was confirmed that the ${}^{238}\rm{U}$ values obtained from the upper stream were consistent with those obtained from the middle stream for the PI(w/GC) 800~$\si{\micro\meter}$ and 100~$\si{\micro\meter}$ parts and the glass cloth,  so that the radiation equilibrium was confirmed.
The finite value of the samples are also shown in converted units of the emission rate~$[\si{\micro\becquerel/cm^2}]$ (Table~\ref{tab:measurment_result2}). 
This unit  is useful to discuss the origin of the $\alpha$-rays which come out of the surface area.
In Table~\ref{tab:measurment_result2}, U/Th amounts of PI(w/GC) 100\,\si{\micro \metre} part are consistent with that of the glass cloth roughly, and thus it was considered that the main U/Th contamination was the contribution by the glass cloth used for the PI(w/GC) 100\,\si{\micro \metre} part.
The glass cloth was taken out of the PI(w/GC) 100\,\si{\micro \metre} layer by dissolving the PI part with a basic aqueous solution.
The reason why the U/Th amounts in the glass cloth were smaller than those in the PI(w/GC) 100\,\si{\micro \metre} part was considered to be that U/Th flowed to the solution when the PI(w/GC) 100\,\si{\micro \metre} part was dissolved.
\par
In NEWAGE, $\gamma$- and $\beta$-ray events are cut with a rejection power of $\sim10^{-5}$ around the threshold, but $\alpha$-ray events cannot be discriminated from nuclear recoils.
The track length of the un-contained $\alpha$-ray and the nucleus around the threshold are both about 1~mm.
Therefore, the remaining background could be $\alpha$-ray events.
The SRIM\cite{SRIM} simulation showed that $\alpha$-rays emitted from the PI(w/GC) 800\,\si{\micro \metre} part can not pass through  the PI(w/GC) 100\,\si{\micro \metre} part on the near side of the detection volume.
Therefore, $\alpha$-rays from the PI(w/GC) 800\,\si{\micro \metre} part and PI(w/GC) 100\,\si{\micro \metre} part on the far side from the detection volume (PI(w/GC) 100\,\si{\micro \metre}~(FS)) are not considered as the background sources.
Therefore, it is required to change the PI(w/GC) 100\,\si{\micro \metre} part on the near side of the detection volume with a low-background material for improved detector sensitivity.
Considering the production risk of changing the thickest layer, we decided to retain the PI(w/GC) 800\,\si{\micro \metre} part in the development of the low-$\alpha$ $\mu$-PIC.
\par 
The goal of this study is to develop a $\mu$-PIC with an $\alpha$-ray emission rate 100 times lower than the standard one, so as to allow NEWAGE to search in the DAMA region~\cite{DAMA_2004} (Figure~\ref{fg:future_limit_SD}).
\par 
The requirements for the new $\mu$-PIC were set as follows.

\begin{itemize}

\item The non-uniformity of the gas gain should be $< 20\%$ in RMS.\\
This value was decided on the basis of a sufficient energy resolution and $\gamma$-ray events' discrimination.
If the non-uniformity of the gas gain is worse than 20~\%, low-energy $\gamma$-ray events could be misidentified as nuclear recoil region of the energy range of interest.

\item The gas gain must be $>1000$ using the argon-ethane gas mixture~(9:1) at 1~atm.\\
Electrons are amplified by a strong electric field near the anode electrode.
Gas gain is defined by the number of electrons amplified by the $\mu$-PIC.
This value is to have  sufficient detection efficiency for nuclear recoil events with the~50 keV threshold under dark-matter search conditions~($\rm{CF}_{4}$ at 0.1~atm).

\item $\alpha$-ray emission rate should be reduced to the level of standard $\mu$-PIC $\times$ 1/100.\\
This value is the background level required to reach the DAMA region.
\end{itemize}

\begin{table}[H]
\centering
\caption{Contamination of U/Th-chain isotopes in the materials of the standard $\mu$-PIC detector components, as measured using the HPGe detector. The uncertainties listed are statistical errors. The upper limits are at 90\% confidence level.}
\scalebox{0.7}{
\begin{tabular}{l|c|c|c}
\hline
Sample & $^{238}\rm{U}$ upper stream [$10^{-6}$ g/g] &$^{238}\rm{U}$ middle stream [$10^{-6}$ g/g] & $^{232}\rm{Th}$ [$10^{-6}$ g/g]\\ \hline \hline
PI(w/GC)~$800\,\si{\micro \metre}$& $0.78\pm0.01$ &$0.76\pm0.01$ &  $3.42\pm0.03$\\
PI(w/GC)~$100\,\si{\micro \metre}$& $0.38\pm0.01$ &$0.39\pm0.01$ &  $1.81\pm0.04$\\
glass cloth(GC) &$0.91\pm0.02$ & $0.84\pm0.03$ & $3.48\pm0.12$\\ 
plating solution(CuSO$_{4}$) & $<0.13$ & $<0.01$ & $<0.06$\\ \hline
GEM & $<0.17$ & $<0.02$ &  $<0.12$\\
\hline
\end{tabular}
}
\label{tab:measurment_result}
\end{table}

\begin{table}[H]
\centering
\caption{Radioactivities of the U/Th-chain isotopes in a sample PI(w/GC)~$100\,\si{\micro \metre}$ insulator sheet. 
These values are recalculated from the results of Table\,\ref{tab:measurment_result}.
The uncertainties listed are statistical errors.}
\scalebox{0.7}{
\begin{tabular}{l|c|c|c}
\hline
Sample & $^{238}\rm{U}$ upper stream $[\si{\micro\becquerel/cm^2}]$&$^{238}\rm{U}$ middle stream $[\si{\micro\becquerel/cm^2}]$&$^{232}\rm{Th}$ $[\si{\micro\becquerel/cm^2}]$\\ \hline \hline
PI(w/GC)~$100\,\si{\micro \metre}$& $66.4\pm2.5$ & $68.5\pm1.5$ &  $102.1\pm2.3$\\
glass cloth(GC)& $71.3\pm1.4$&$64.5\pm0.8$ &$86.8\pm1.1$\\ 
\hline
\end{tabular}
}
\label{tab:measurment_result2}
\end{table}

\subsection{Material selection}
\label{subsec:material_selection}
In order to make a new $\mu$-PIC whose $\alpha$-ray emission would be less than 1/100 of the standard $\mu$-PIC, it was necessary to search for a material containing 1/100 or less U/Th than the PI(w/GC)$100\,\si{\micro \metre}$ part did.
A material consisting of a PI layer without glass cloth(w/o GC) and with an epoxy layer was chosen as a new candidate material as illustrated in Figure~\ref{fig:cs_uPIC}.
It has a layered structure with $80\,\si{\micro \metre}$ thickness, with a PI(w/o GC) part of $5~\si{\micro \metre}$ and an epoxy part of $75\,\si{\micro \metre}$.
The amount of U/Th contamination in this new material was measured with an HPGe detector in the Kamioka underground laboratory.
This HPGe detector was different from the one used for the standard $\mu$-PIC measurements. 
Details of the detector system have been reported by Abe et al.~\cite{HPGe}.
The measurement results using the HPGe detector are summarized in Table~\ref{tab:LA_measurment_result}.
We assumed a radioactive equilibrium and took the value of $^{238}$U middle stream.
The new material had less than $1/100$ U/Th contamination than the PI(w/GC)$100~\si{\micro \metre}$ layer used in the standard $\mu$-PIC.

\begin{table}[H]
\centering
\caption{$^{238}$U and $^{232}$Th measurement results using the HPGe detector. The uncertainties listed are statistical errors. The upper limits are 90\% confidence level.}
\scalebox{0.7}{
\begin{tabular}{l|c|c|c}
\hline
Sample & $^{238}\rm{U}$ upper stream [$10^{-6}$ g/g]&$^{238}\rm{U}$ middle stream [$10^{-6}$ g/g] &$^{232}\rm{Th}$ [$10^{-6}$ g/g]\\ \hline \hline
PI(w/GC) 100\,\si{\micro \metre}& $0.38\pm0.01$ &$0.39\pm0.01$& $1.81\pm0.04$\\
PI(w/o GC)+epoxy & $<2.86 \times 10^{-2}$&$<2.98\times10^{-3}$ & $<6.77\times10^{-3}$\\
\hline
\end{tabular}
}
\label{tab:LA_measurment_result}
\end{table}

\subsection{Structural check of the low-$\alpha$ $\mu$-PIC}
\begin{figure}[H]
\centering
\includegraphics[width=13cm]{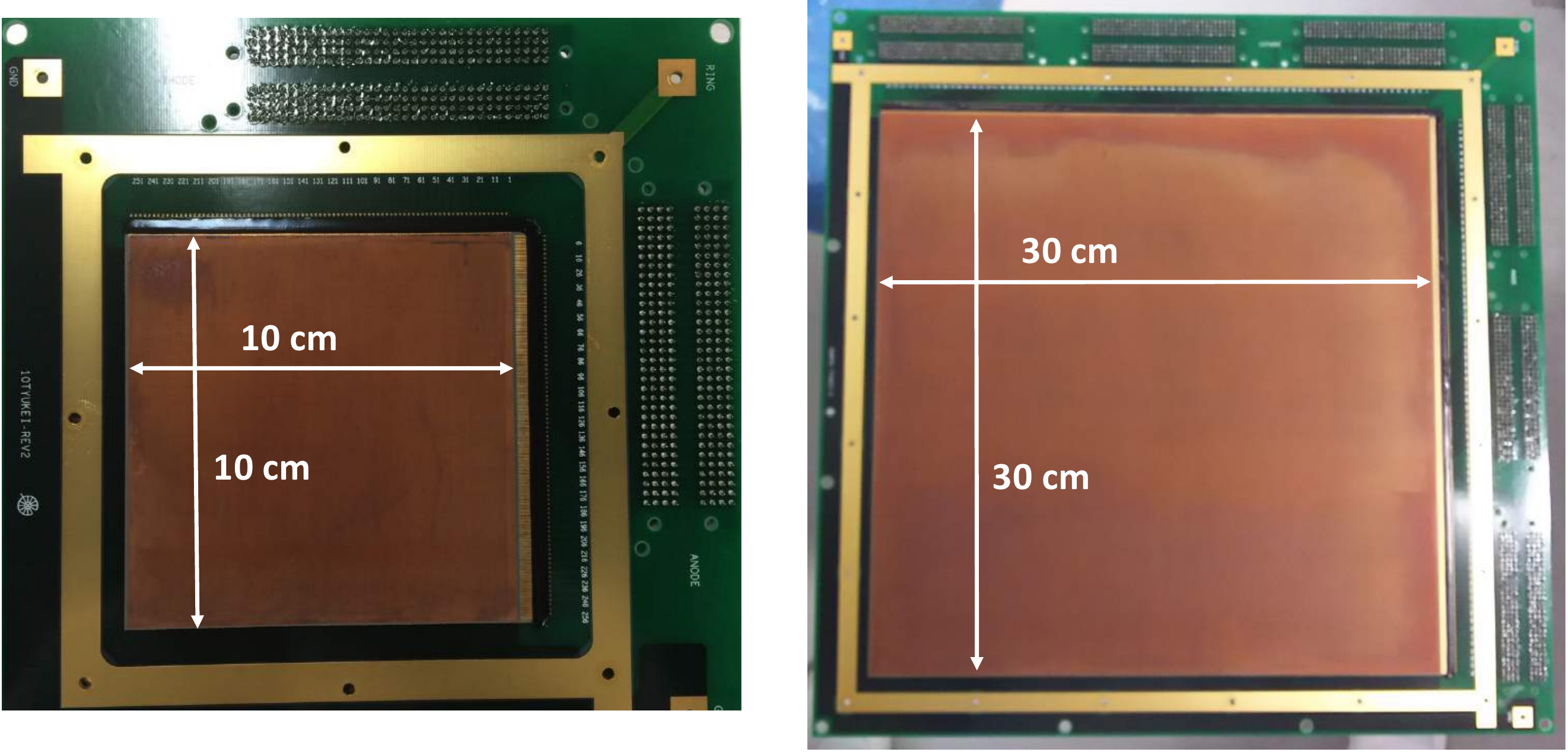}
\caption{Pictures of a $10\times10~\rm{cm^{2}}$ low-$\alpha$ $\mu$-PIC prototype (left) and $30\times30~\rm{cm^{2}}$ low-$\alpha$ $\mu$-PIC (right).}
\label{fig:LAuPIC_gaikan}
\end{figure}

\begin{figure}[H]
\centering
\includegraphics[width=120 mm]{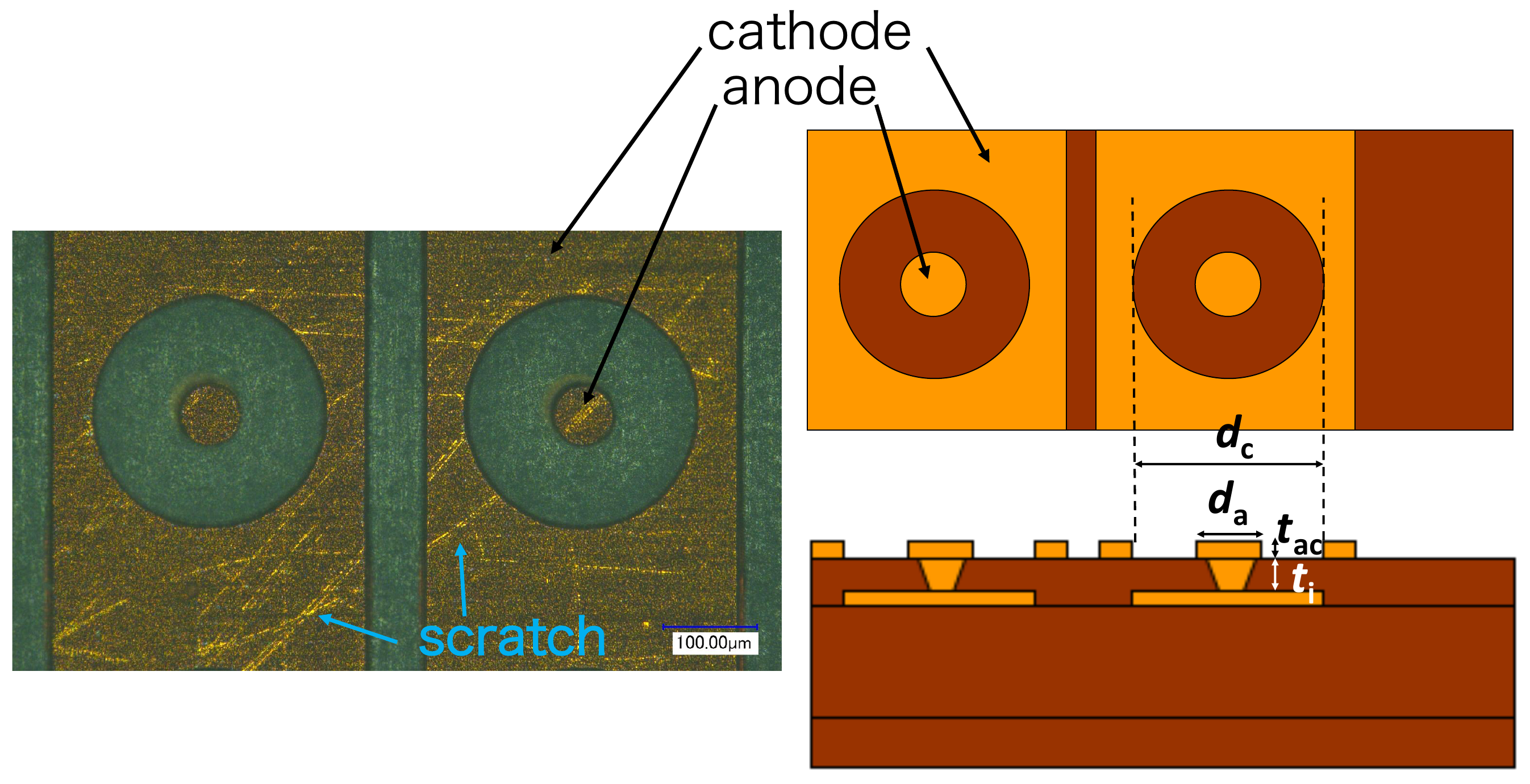}
\caption{Left panel is the electrodes of the low-$\alpha$ $\mu$-PIC observed with the digital microscope.
Right panel is the schematic design for the parameters relevant to the detector performance.}
\label{fg:uPIC_param}
\end{figure}

A new $\mu$-PIC named ``low-$\alpha$ $\mu$-PIC'' was manufactured by replacing the PI(w/GC)100\,\si{\micro \metre} parts with new material discussed in Section~\ref{subsec:material_selection}.
The cross-section view of the low-$\alpha$ $\mu$-PIC is shown in the right panel of Figure~\ref{fig:cs_uPIC}.
The sizes of the produced low-$\alpha$ $\mu$-PIC were $10\times10~\rm{cm^{2}}$ and $30\times30~\rm{cm^{2}}$.
The $10\times10~\rm{cm^{2}}$ low-$\alpha$ $\mu$-PICs were made as prototypes and the $30\times30~\rm{cm^{2}}$ low-$\alpha$ $\mu$-PICs were made to be used for the dark matter search.
Figure~\ref{fig:LAuPIC_gaikan} (left) and (right) show the appearance of the $10\times10~\rm{cm^{2}}$ and $30\times30~\rm{cm^{2}}$ low-$\alpha$ $\mu$-PICs, respectively.
In their production, no substantial problem were found due to the material change.
The principle of the gas amplification of the $\mu$-PIC is basically the same as that of a proportional counter.
The electron is amplified by the strong electric field around the anode electrode.
Therefore, understanding the structure around the anode electrode is important.
The structures of the electrodes of the manufactured low-$\alpha$ $\mu$-PICs were measured using a digital microscope (KEYENCE VHX-2000).
The picture obtained with the digital microscope is shown in the left panel of Figure~\ref{fg:uPIC_param} and the parameters relevant to the detector performance are defined in the right panel of Figure~\ref{fg:uPIC_param}.
Although scratches on the cathode surface were confirmed, such scratches were also seen on the standard $\mu$-PIC and were known not to cause any problem for the operation.
The diameters of the anode electrodes~($d_{\rm{a}}$), the diameters of the cathode openings~($d_{\rm{c}}$), and the height of the electrodes~($t_{\rm{ac}}$) were measured.
The measurement results are listed in Table~\ref{tab:mesurment_result_uPICparam}.
Since the thickness of the insulator ($t_{\rm{i}}$ in Figure~\ref{fg:uPIC_param}) could not be measured for the final products, the design values are shown.
The structures in the four corners~(about 4$~\si{mm}$ inside from the edge) and the center of the detectors were measured, and the average values and their standard deviations at these five locations are listed in Table~\ref{tab:mesurment_result_uPICparam}.
The measurement results of the $d_{\rm{c}}$ and $t_{\rm{ac}}$ of low-$\alpha$ $\mu$-PICs were smaller than the design values.
These are due to material changes and can be improved by taking expansion and contraction during the manufacture into consideration.

\begin{table}[H] 
\centering
\caption{Measurement results and the design value of each parameter of the $10\times10$~\si{\cm^{2}} low-$\alpha$ $\mu$-PIC and the $30\times30$~\si{\cm^{2}} low-$\alpha$ $\mu$-PIC. The errors are the standard deviation.
The parameters in the table are defined in Figure~\ref{fg:uPIC_param}.
}
\scalebox{0.8}{
\begin{tabular}{l|c|c|c}
\hline
Parameter &$10\times10$~\si{\cm^{2}} low-$\alpha$ $\mu$-PIC & $30\times30$~\si{\cm^{2}} low-$\alpha$ $\mu$-PIC &Design value \\ \hline \hline
$d_{\rm{a}}$~[\,\si{\um}]&$64.4\pm2.8$&$62.9\pm2.5$& 60 \\ 
$d_{\rm{c}}$~[\,\si{\um}]&$240.0\pm3.0$&$242.3\pm2.3$&250 \\ 
$t_{\rm{ac}}$~[\,\si{\um}]&$15.4\pm1.1$&$14.1\pm1.4$&20\\ 
$t_{\rm{i}}$~[\,\si{\um}]&NA&NA&80\\
\hline
\end{tabular}
}
\label{tab:mesurment_result_uPICparam}
\end{table}

\section{Performance of Low-$\alpha$ $\mu$-PIC}

\subsection{Setup}
The performance of the low-$\alpha$ $\mu$-PICs was measured in a test chamber.
This measurement was performed with a gas flow of argon-ethane mixture~(9:1).
The position dependence and the anode voltage dependence of the gas gains of the low-$\alpha$ $\mu$-PICs were measured.
\par
\begin{figure}[t]
\centering
\includegraphics[width=100 mm]{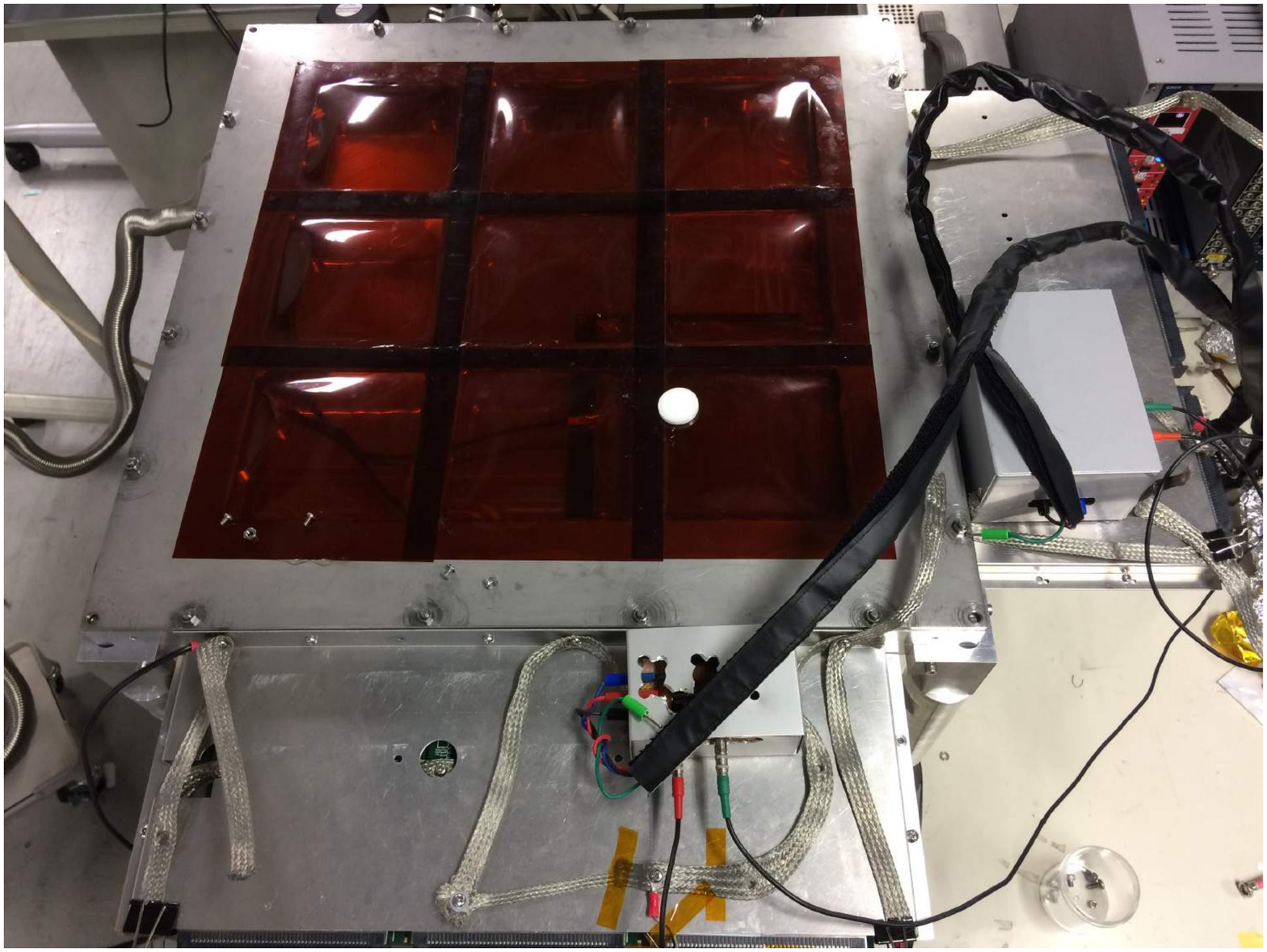}
\caption{Outer view of a chamber used to evaluate a performance of the low-$\alpha$ $\mu$-PICs.}
\label{fg:check_detector}
\end{figure}
\begin{figure}[t]
	 \centerline{\includegraphics[width=100mm]{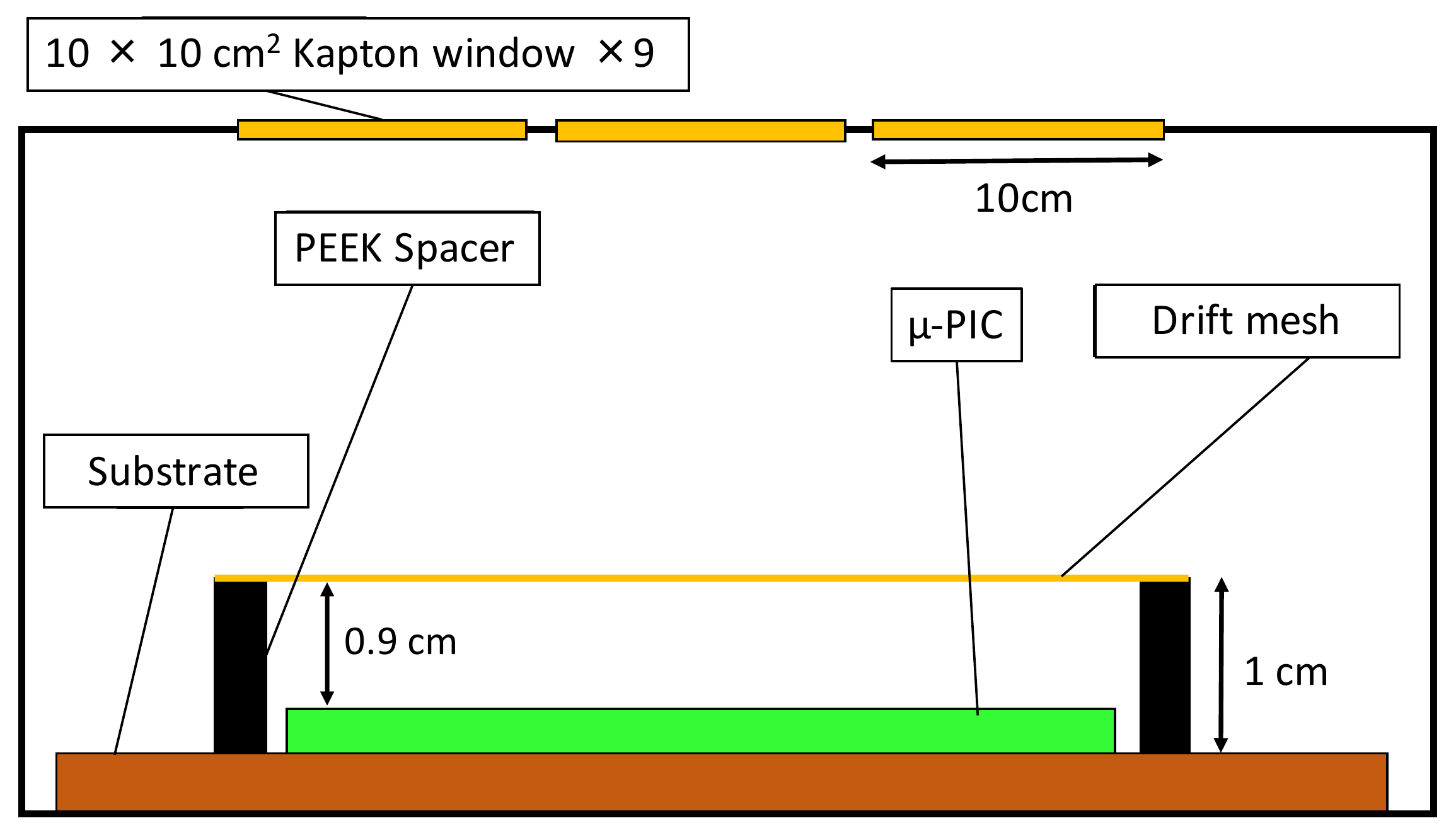}}
  \caption{Schematic cross section of the detector used for a performance check of the low-$\alpha$ $\mu$-PICs. }
  \label{fig:detector_ponchi}
\end{figure}

The outer view of the test chamber and a schematic crosssection of the setup used for the performance tests are shown in Figures~\ref{fg:check_detector} and \ref{fig:detector_ponchi}, respectively.
The test chamber was made of aluminum.
There were nine $10\times10$~\si{\cm^{2}} kapton windows with a thickness of 125$~\si{\um}$.
The drift mesh was made of SUS304 with a wire diameter of 20$~\si{\um}$ and a mesh pitch of 68$~\si{\um}$, and the opening ratio was 59.7\%. 
The detector was set in the chamber with a gas flow of argon-ethane mixture~(9:1).
The gas flow rate was 30~\si{\cm^{3}}/min.
During the measurement, gas flow was performed using a gas blender (SECB-2) which also controlled the gas flow rate.
A drift voltage of -500V was supplied to the drift mesh that formed a drift field of 0.55~kV/cm in a drift length of 0.9~$\si{\cm}$.

\par
The gas gains were measured using 5.9~keV X-rays from a radioactive source of $^{55}\rm{Fe}$.
The rate of $^{55}\rm{Fe}$ source was 1.0~MBq.
Thirty-two strips of the anode electrodes were connected and the charges from them were amplified by a charge-sensitive amplifier and used as a data acquisition trigger. 
The anode signal was used only for the data acquisition trigger and was not recorded.
Thirty-two strips of cathode electrodes were connected, and their charges were amplified by a charge-sensitive amplifier and recorded with a waveform digitizer. 
CREMAT CR-110 was used for the anode amplifier, and CREMAT CR-110 and LF356N were used for the cathode amplifier.
The gain of the anode and cathode amplifiers was $-0.7$~V/pC and $-4.0$~V/pC, respectively.
The shaping time of the cathode amplifier was about 1~\si{\us}.
The waveform data were stored using a flash ADC (GNV-240G) at 500~MHz sampling rate.
The dynamic range and resolution of the voltage were $0-1$~V with 8 bits and the sampling depth was $8168$.
It was confirmed by a calibration with the test pulse. 
In addition, it was confirmed that no saturation occurred until 0.16~pC by checking the signal waveforms.
\par
\begin{figure}[ht]
	 \centerline{\includegraphics[width=80mm]{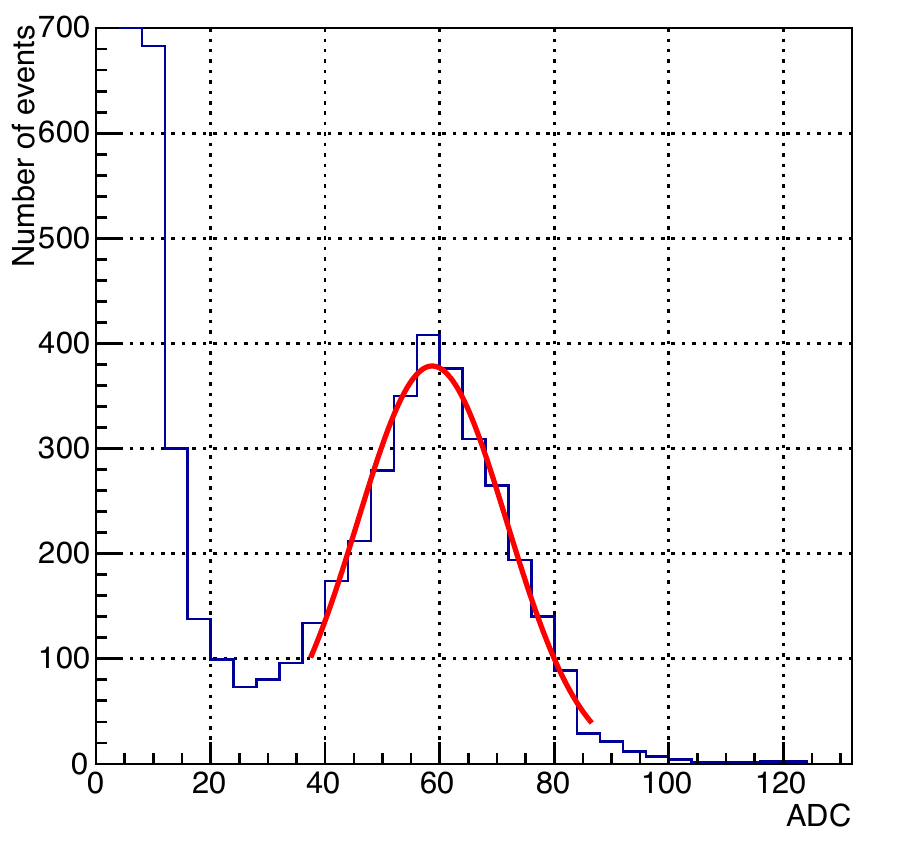}}
	  \caption{Energy spectrum obtained with ${}^{55}\rm{Fe}$ X-rays measured by the $30\times30$~\si{\cm^{2}} low-$\alpha$ $\mu$-PIC.
	  Anode voltage was $520$~V. 
	  The energy spectrum obtained with the $10\times10$~\si{\cm^{2}} low-$\alpha$ $\mu$-PIC is of similar shape.
	 }
  \label{fig:energy_spectrum}
\end{figure}

An energy spectrum of ${}^{55}\rm{Fe}$ X-rays measured by the $30\times30$~\si{\cm^{2}} low-$\alpha$ $\mu$-PIC is shown in Figure~\ref{fig:energy_spectrum}.
The anode voltage was $520$~V.
The main peak at 5.9~keV was used in the gas gain measurement.
The gas gains were determined by fitting the main peak with Gaussian distribution for each measurement and the mean values were used.
The gas gain was defined as follows

\begin{equation}
G_{{\rm gas}} = \frac{W_{{\rm Ar}} \times (a\times{\rm ADC}-b)}{(5.9~{\rm keV})\times e^{-}}
\label{eq:gas_gain}
\end{equation}
where $\rm{W_{Ar}}$ is the W value of the gas mixture~(26 eV)\cite{W_value}, $e^{-}$ is the elementary charge, and $a$ and $b$ are the calibration factors of the cathode amplifier.
The gain factor $a$ and the offset $b$ were measured by inputting test pulses and their values were $a=1.12\times10^{-3}$~pC/ADC and $b=5.71\times10^{-3}$~pC.

\subsection{The measurement results}
\begin{figure}[H]
	 \centerline{\includegraphics[width=150mm]{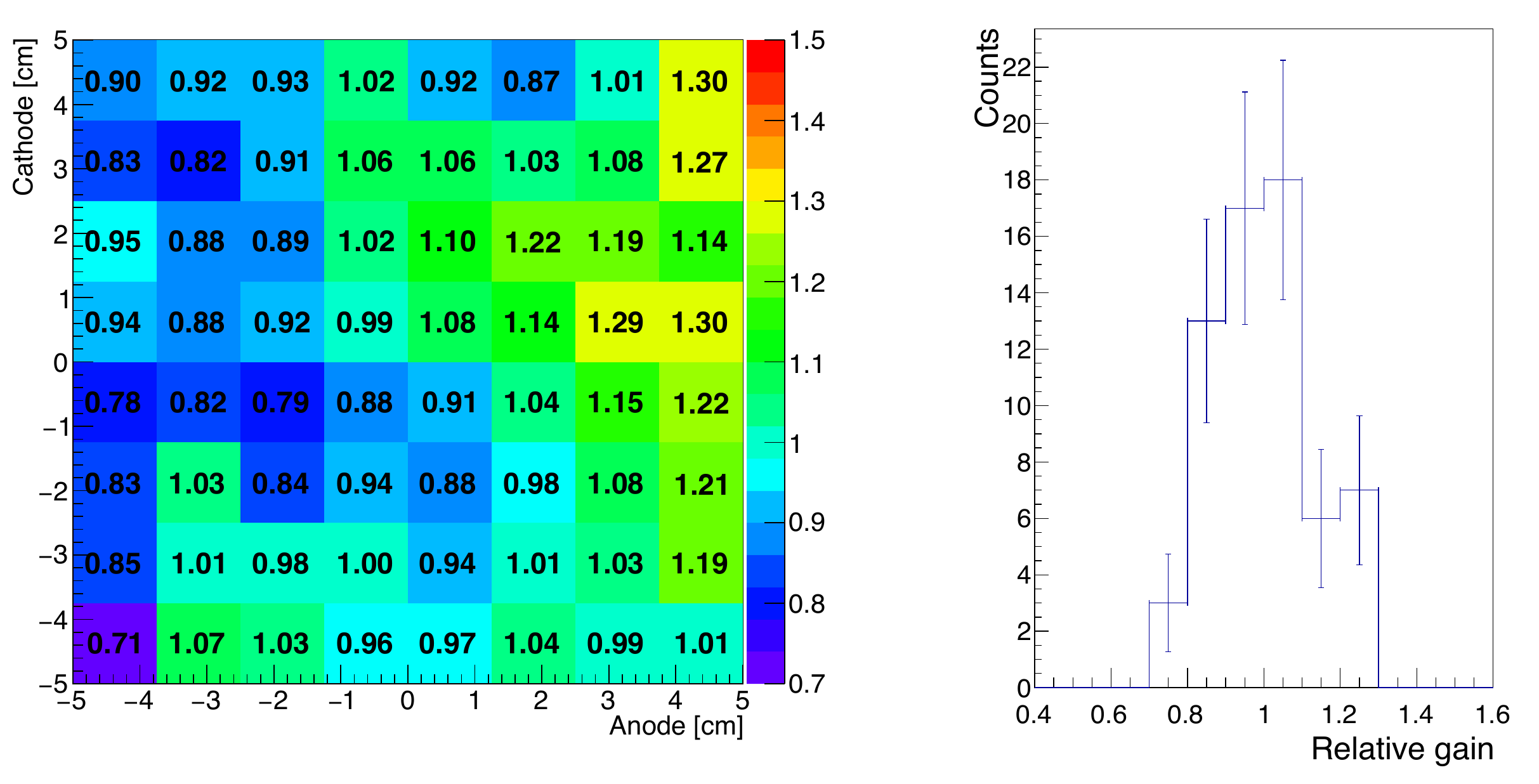}}
  \caption{Left panel is the non-uniformity of gas gain of a 10 $\times$10~\si{\cm^{2}} low-$\alpha$ $\mu$-PIC. 
  Anode voltage was $530$~V. 
  Numbers shown in the left panel are relative gains.
  Right panel is the histogram of relative gains.
  Statistical errors are shown.
  }
  \label{fig:10LA_gainmap}
\end{figure}

The gas gains were measured for a prototype $10\times10$~\si{\cm^{2}} low-$\alpha$ $\mu$-PIC (SN 160115-2) and a $30\times30$~\si{\cm^{2}} low-$\alpha$ $\mu$-PIC (SN 161130-5).
The measurement of the non-uniformity of the gas gain of the $10\times10$~\si{\cm^{2}} low-$\alpha$ $\mu$-PIC was carried out at an anode voltage of $530$~V.
The measurement result of the non-uniformity of the $10\times10$~\si{\cm^{2}} low-$\alpha$ $\mu$-PICs is shown in Figure~\ref{fig:10LA_gainmap}.
The numbers in the histogram represent relative gains.
The measured gains were normalized by the average value so that the relative gains can be shown.
The non-uniformity of the gas gain of the $10\times10$~\si{\cm^{2}} low-$\alpha$ $\mu$-PIC was $13\%$ RMS.

\par
\begin{figure}[H]
	 \centerline{\includegraphics[width=150mm]{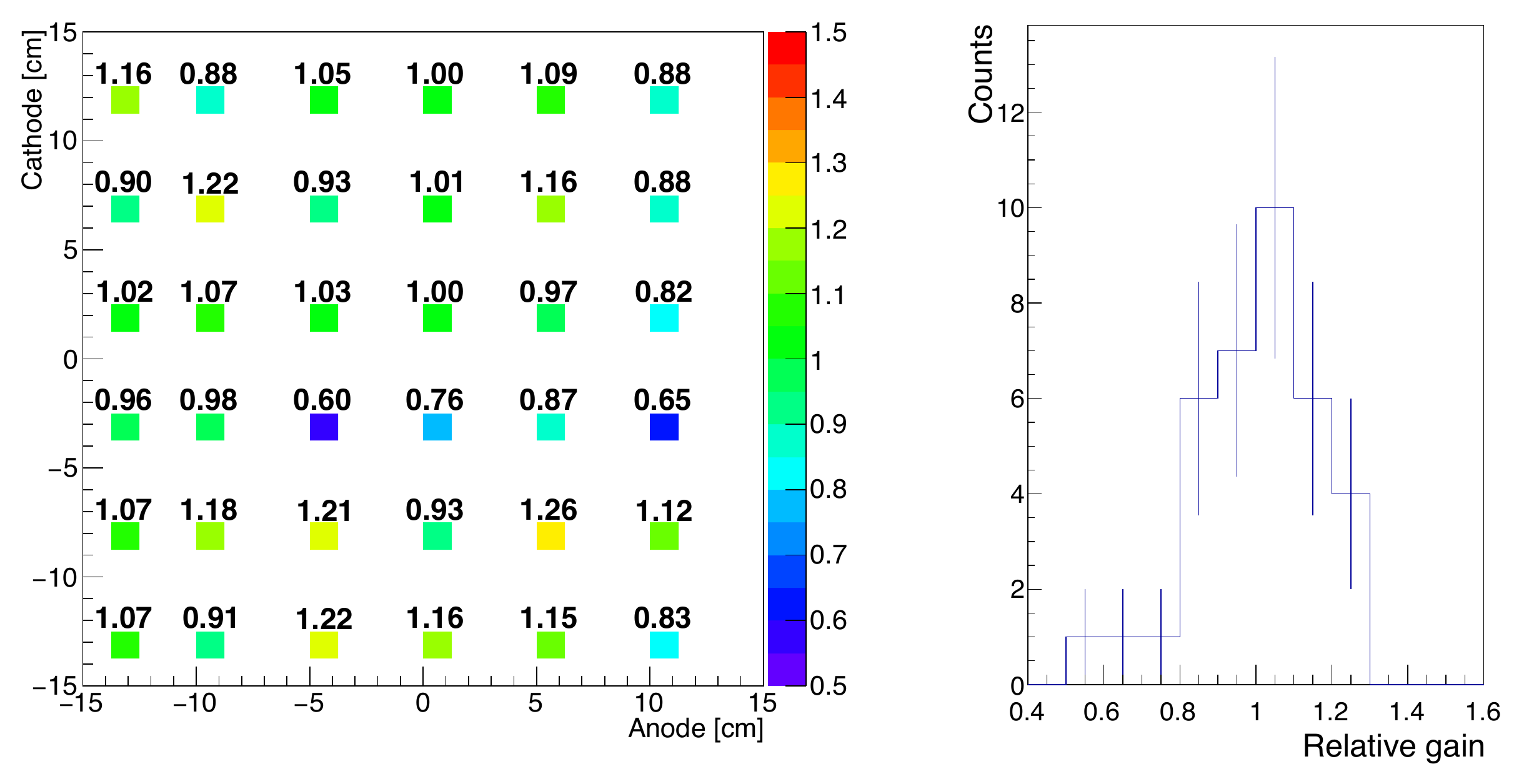}}
  \caption{Left panel is the position dependence of the gas gain of the $30\times30$~\si{\cm^{2}} low-$\alpha$ $\mu$-PIC. 
  Numbers shown in the left panel are relative gains.
  Right panel is the histogram of the relative gains.
  Statistical errors are shown.}
  \label{fig:30LA_gainmap}
\end{figure}
\begin{figure}[H]
	 \centerline{\includegraphics[width=100mm]{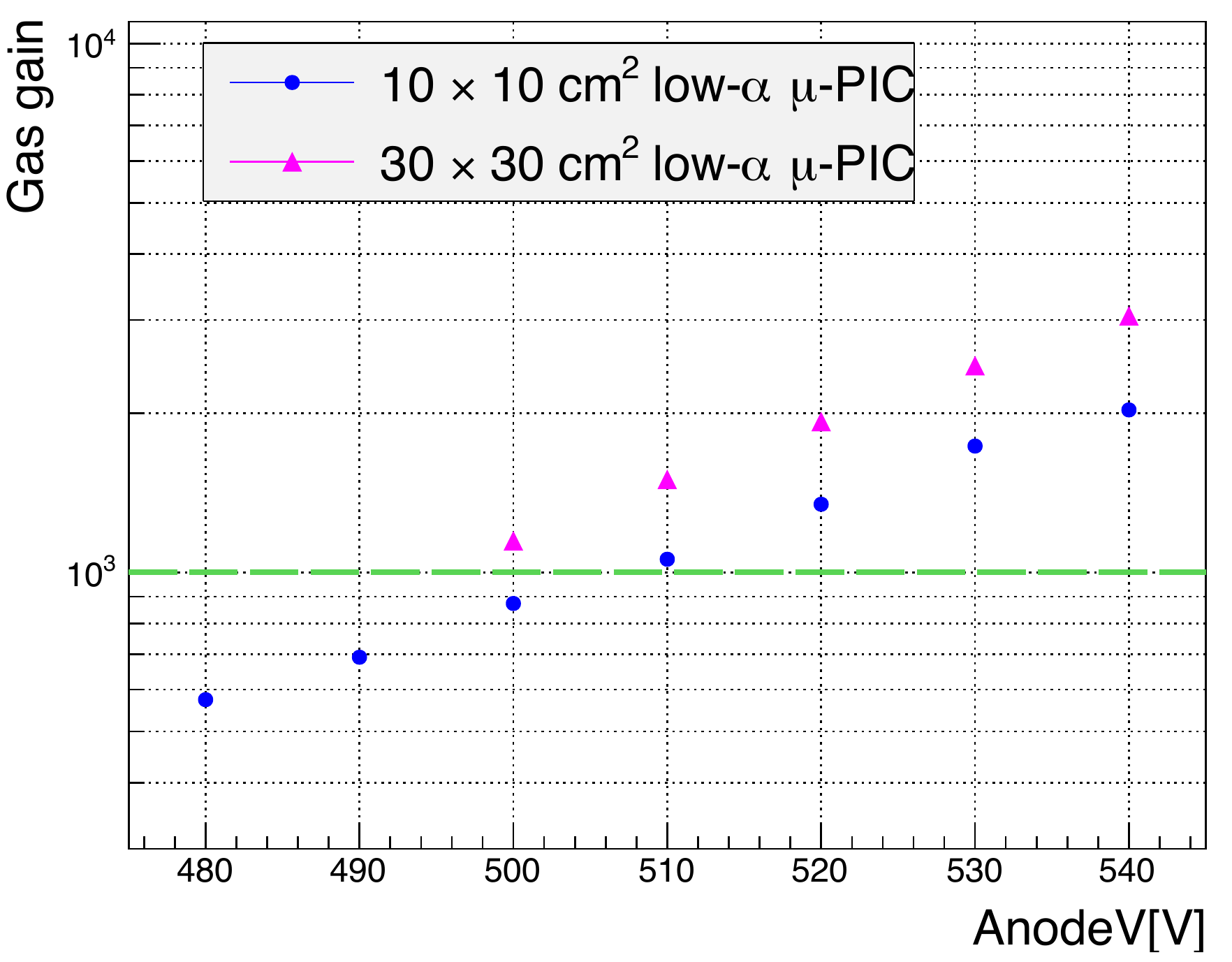}}
	  \caption{Gas gains as a function of the voltage supplied to the anode electrodes of the $10\times10$~\si{\cm^{2}} and $30\times30$~\si{\cm^{2}} low-$\alpha$ $\mu$-PIC. 
	  The green dash line represents the requirement of the gas gain.
	  The difference in the gain between $10\times10$~\si{\cm^{2}} and $30\times30$~\si{\cm^{2}} low-$\alpha$ $\mu$-PICs is within the expected variation from the production considering the experience with the standard $\mu$-PICs.}
  \label{fig:gc1}
\end{figure}

The $30\times30$~\si{\cm^{2}} low-$\alpha$ $\mu$-PIC was sampled at $36~(6\times6)$ points, and the gas gains were measured using the $1.25\times1.25$~\si{\cm^{2}} region (the intersection of 32 anode strips $\times$ 32 cathode strips) as a representative area in each sampling area of $5\times5$~\si{\cm^{2}}.
The measurement results of the $30\times30$~\si{\cm^{2}} low-$\alpha$ $\mu$-PIC are shown in Figure~\ref{fig:30LA_gainmap}.
The squares represent the measurement positions and their sizes.
The measurement of the position dependence of the $30\times30$~\si{\cm^{2}} low-$\alpha$ $\mu$-PIC was carried out at an anode voltage of between 520 and 540~V.
For the measurement points at a voltage different from 540~V, the value was corrected to the one corresponding to the 540~V gas gain based on the gain curve shown in Figure~\ref{fig:gc1}.
The non-uniformity of the gas gain of the $30\times30$~\si{\cm^{2}} low-$\alpha$ $\mu$-PIC was $16\%$~($\equiv~\sigma_{\rm{LA30cm}}$) RMS.
The gain variation due to the unmeasured 83\% of the the total surface area of the $30\times30$~\si{\cm^{2}} low-$\alpha$ $\mu$-PIC was considered.
The variation within a $5\times5$~\si{\cm^{2}} area was estimated from the measurement result of the $10\times10$\,\si{\cm^{2}} low-$\alpha$ $\mu$-PIC.
The $10\times10$~\si{\cm^{2}} low-$\alpha$ $\mu$-PIC was divided into four parts, the non-uniformity of the gas gain was obtained for each, and the average value ($\equiv \sigma_{\rm{LA5cm}}$) of $10\%$ RMS was obtained.
From these measurements, the non-uniformity of the gas gain of the whole $30\times30$~\si{\cm^{2}} low-$\alpha$ $\mu$-PIC~($\equiv~\sigma_{\rm{all\,LA30cm}}$) was known to be $\sigma_{\rm{all\,LA30cm}} = \sqrt{\sigma_{\rm{LA5cm}}^2 + \sigma_{\rm{LA30cm}}^2} = 19\%$ at the RMS.
This value satisfied the 20\% requirement.

The anode voltage dependencies of the gas gain for the $10\times10$~\si{\cm^{2}} and $30\times30$~\si{\cm^{2}} low-$\alpha$ $\mu$-PICs are shown in Figure~\ref{fig:gc1}.
The gas gains were corrected for an averaged gas gain of the position dependence (relative gas gain~$=1$).
The measurement points of the $10\times10$~\si{\cm^{2}} and $30\times30$~\si{\cm^{2}} low-$\alpha$ $\mu$-PICs were the region from 0 to $1.25$~\si{\cm} of the anode and from 0 to 1.25~\si{\cm} of the cathode (Figure~\ref{fig:10LA_gainmap}), and the region from 5 to 6.25~\si{\cm} of the anode and from $-8.75$ to $-7.50$~\si{\cm} of the cathode (Figure~\ref{fig:30LA_gainmap}), respectively.
The range of applied voltage was 480~V to 540~V for the $10\times10$~\si{\cm^{2}} and 500~V to 540~V for the $30\times30$~\si{\cm^{2}}.
The higher limit was determined by the discharges, and the lower limit was determined by the electric noise.
When the voltage is higher than 540~V, the discharge is more than the events from the source and the energy spectrum is not visible.
The requirement of the gas gain with argon-ethane gas mixture~(9:1) was $1.0\times10^{3}$.
From Figure~\ref{fig:gc1}, both $10\times10$~\si{\cm^{2}} and $30\times30$~\si{\cm^{2}} low-$\alpha$ $\mu$-PICs achieved average gas gains of 1000 at 510~V and were found to satisfy the requirement.
The difference in the gain between $10\times10$~\si{\cm^{2}} and $30\times30$~\si{\cm^{2}} low-$\alpha$ $\mu$-PICs is within the expected variation from the production considering the experience with the standard $\mu$-PICs.
There are individual differences in the gas gain depending on the state of the electrode of  $\mu$-PIC (height, diameter, shape, etc.).

\subsection{Surface $\alpha$-ray emission rate}
The $\alpha$-ray emission rates of the standard and low-$\alpha$ $\mu$-PICs were measured by an $\alpha$-ray counter, specifically an Ultra-Lo 1800 made by XIA LLC~\cite{UltraLo}.
This emission rate is the sum of the surface and bulk of the sample.
The achievable background rate is $\sim 10^{-5}$ ${\rm \alpha/\si{\cm^{2}}/h}$.
The analysis method of the $\alpha$-rays, especially separation between $\alpha$-rays from bulk and from surface of samples using Ultra-Lo~1800 was established by the XMASS group~\cite{UltraLo2}.
The $\alpha$-rays coming from $\alpha$ decay on the sample surface make a peak in the energy spectrum and the $\alpha$-rays coming from $\alpha$ decay in the bulk of the sample make a continuous spectrum.
\par
\begin{figure}[t]
	 \centerline{\includegraphics[width=120mm]{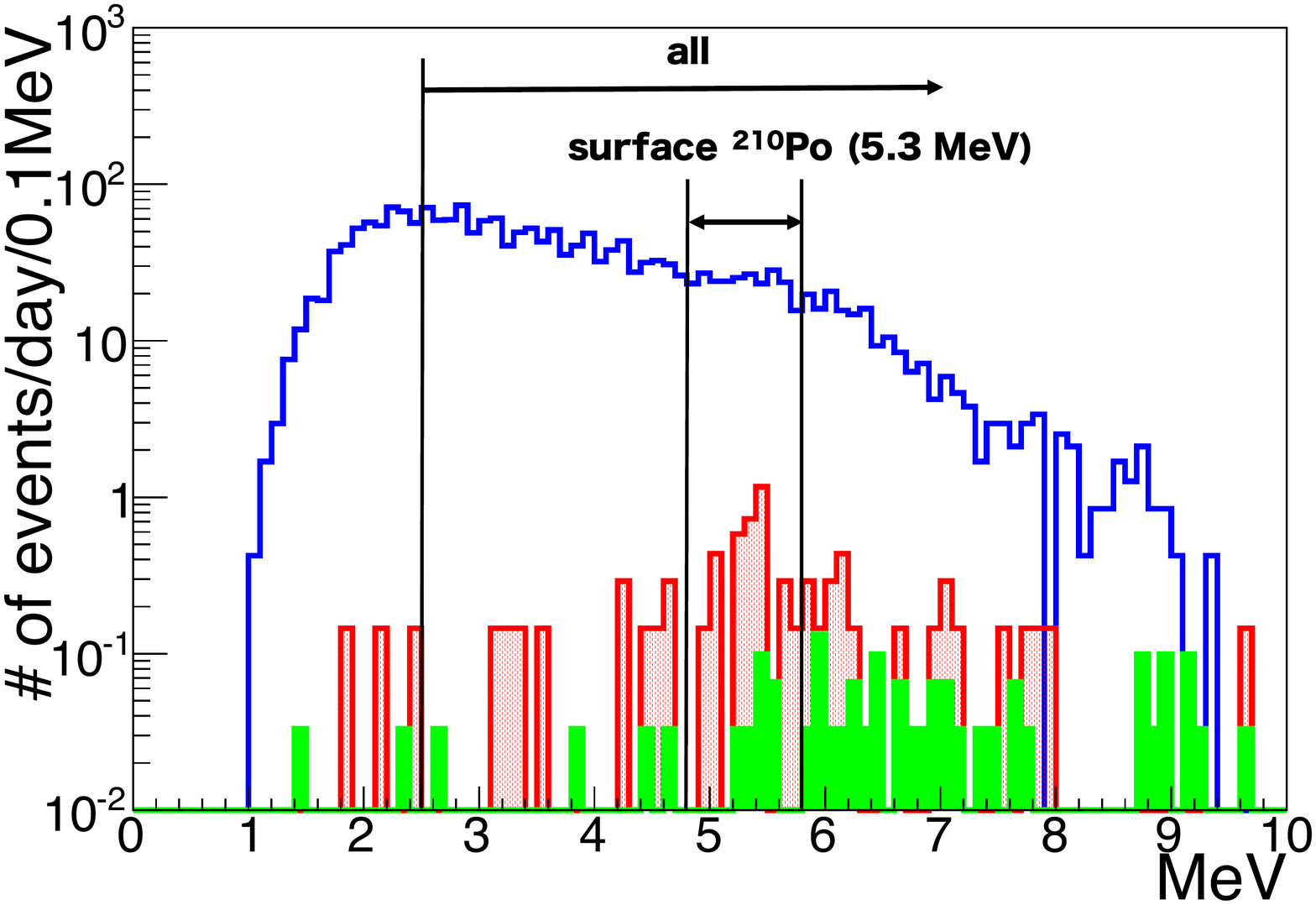}}
	  \caption{The blue, red and green histograms show the measurements of the standard $\mu$-PIC, the low-$\alpha$ $\mu$-PIC and the silicon wafer, respectively.
	  Because the silicon wafer sample is clean, it shows the spectrum of the background of the $\alpha$-ray counter.
	  }
  \label{fig:enecom}
\end{figure}

The energy spectra taken by the Ultra-Lo 1800 are shown in Figure~\ref{fig:enecom}.
The sample sizes of the standard $\mu$-PIC and the low-$\alpha$ $\mu$-PIC were 25 sheets of $5\times5$~\si{\cm^{2}} square and a $30\times30$~\si{\cm^{2}} square sheet, respectively.
The background was measured using a silicon wafer as a blank sample.
It is known that the silicon wafer is clean enough to be used for the background measurement.
A large continuous component without any clear peak was seen in the energy spectrum of the standard $\mu$-PIC, while a clear 5.3~MeV was seen in the energy spectrum of the low-$\alpha$ $\mu$-PIC.
The peak seen in the low-$\alpha$ $\mu$-PIC is thought to mainly due to the $^{210}{\rm Po}$ decay on the sample surface.

\par
The measurement results are summarized in Table~\ref{tab:LA_measurment_result_UltraLo}.
The events with an energy larger than 2.5~MeV were used for this analysis.
We found that the total $\alpha$-ray emission rate (2.5~MeV $<E$ in Table~\ref{tab:LA_measurment_result_UltraLo}) of the low-$\alpha$ $\mu$-PIC was reduced to less than $1/100$ compared to that of the standard $\mu$-PIC and the requirement of the low-$\alpha$ $\mu$-PIC was achieved.
\par
The largest systematic error was thought to be the difference of the sample size.
Since the size of the standard $\mu$-PIC available as the sample was smaller than the active area of the $\alpha$-ray counter (circle with a diameter of $30~\si{\cm}$), $\alpha$-rays originating from the glass cloth could come into the detection volume from the edges of each small pieces. 
On the other hand, the low-$\alpha$ $\mu$-PIC sheet was larger than the active area.
Therefore, no edge effect was expected.
This edge effect was estimated by comparing $5\times5$~\si{\cm^{2}} square and a $30\times30$~\si{\cm^{2}} square sheet samples of low-$\alpha$ $\mu$-PIC.
The low-$\alpha$ $\mu$-PICs contain the glass cloth in the PI(w/GC)~$800\,\si{\micro \metre}$ layer as shown in the right panel of Figure~\ref{fig:cs_uPIC}.
The edge effect in the region of  $2.5$~MeV$<E$ and $4.8$~MeV$<E<5.8$~MeV were $2.0\times 10^{-2}$ and $3.0\times 10^{-3}$~${\rm \alpha/\si{\cm^{2}}/h}$ , respectively.
These values were treated as systematic errors.
\par
Although our goal was achieved, the $\alpha$-ray emission rate of the low-$\alpha$ $\mu$-PIC was observed finite values for the peak and total emission rate.
The emission rate corresponding to the peak component ($4.8$~MeV$<E<5.8$~MeV) of the low-$\alpha$ $\mu$-PIC was $(2.1\pm0.5)\times10^{-4}$~${\rm \alpha/\si{\cm^{2}}/h}$.
This level of radioactivity can occur if the material was placed in an atmosphere with a typical radon concentration for several days by considering a mechanism discussed in Ref.\cite{Rn}.
Therefore, in order to make a low-$\alpha$ $\mu$-PIC with the surface $\alpha$-ray rate of $10^{-4}$ or less, the detector should not be exposed in the air after the final production and cleaning.
The emission rate corresponding to the total energy region of the low-$\alpha$ $\mu$-PIC was $(5.5\pm0.7)\times10^{-4}$~${\rm \alpha/\si{\cm^{2}}/h}$.
This value corresponds to a radioactivity of $\sim100$~mBq/kg.
This is the estimate of 10~mBq/kg order.
This total radioactivity can also be estimated from the HPGe measurement results shown in Table~\ref{tab:LA_measurment_result}.
The upper limits of $^{238}{\rm U}$ upper, $^{238}{\rm U}$ middle, and $^{232}{\rm Th}$ stream were calculated to be 1065, 147, and 164~mBq/kg, respectively. 
The $\alpha$-ray measurement indicates the contamination of $^{238}{\rm U}$ upper, $^{238}{\rm U}$ middle, and $^{232}{\rm Th}$ stream are less than the upper limit set by HPGe measurement.
Since any increase of the radioactive isotope was not observed from the materials, the manufactured low-$\alpha$ $\mu$-PIC detector achieved the required background level.

\begin{table}[H]
\centering
\caption{The $\alpha$-ray emission rate from the samples of the standard and low-$\alpha$ $\mu$-PIC detectors. 
The unit is ${\rm \alpha/\si{\cm^{2}}/h}$.
Emissivities of $\alpha$-ray events coming from $^{210}{\rm Po}$ decay (5.3~MeV) on the sample surface is distributed in $4.8<E<5.8$~MeV.
}
\scalebox{0.8}{
\begin{tabular}{l|c|c}
\hline
Sample  & 2.5~MeV$<E$ &4.8~MeV$<E<5.8$ MeV\\ \hline \hline
Standard $\mu$-PIC & $(1.27\pm0.02(stat.)\pm0.2(sys.))\times 10^{-1}$&$(1.62\pm0.07(stat.)\pm0.3(sys.))\times 10^{-2}$\\
Low-$\alpha$ $\mu$-PIC& $(5.5\pm0.7(stat.))\times10^{-4}$&$(2.1 \pm 0.5(stat.))\times 10^{-4}$\\
\hline
\end{tabular}
}
\label{tab:LA_measurment_result_UltraLo}
\end{table}

\section{Conclusions}
Direction-sensitive methods could provide strong evidence for the direct detection of WIMPs.
NEWAGE, one of the direction-sensitive dark-matter search experiments, suffered from the radioactive background from its readout device, $\mu$-PIC.
A new readout device, low-$\alpha$ $\mu$-PIC, was developed to improve the dark matter sensitivity by more than a factor of 50.  
This low-$\alpha$ $\mu$-PIC was developed using a material with less radioactive contamination by about a factor of 100 compared with the standard $\mu$-PIC.
Low-$\alpha$ $\mu$-PICs with sizes of $10\times10$ $\rm{cm^2}$ and  $30\times30$ $\rm{cm^2}$ were produced, and their performances were studied in terms of the gas gain and its uniformity.
The low-$\alpha$ $\mu$-PICs achieved the required gas gain ($1.0\times10^{3}$).
The position dependence of the gas gain of the $30\times30~\rm{cm^2}$  low-$\alpha$ $\mu$-PIC satisfied the requirement~(20$\%$ in RMS).  
An $\alpha$-ray emission measurement with Ultra-Lo 1800 confirmed that the surface $\alpha$-ray rate reduced by a factor of 100.
The low-$\alpha$ $\mu$-PIC developed in this work is expected to improve the sensitivity of NEWAGE by about a factor of 50.

\section*{Acknowledgments}
We gratefully acknowledge the cooperation of Kamioka Mining and Smelting Company.
We thank the XMASS collaboration and Yoshizumi Inoue for their help on the low-background measurement technologies.
This work was supported by the Japanese Ministry of Education, Culture, Sports, Science and Technology, Grant-in-Aid for Scientific Research, ICRR Joint-Usage, JSPS KAKENHI Grant Numbers 16H02189, 26104004, 26104005, 26104009, 19H05805, 19H05806, and 19H05808.




\end{document}